\documentstyle[12pt]{article}
\newcommand{\pl}{\partial}
\begin{document}

\begin{center}
{\large\bf q-oscillators, (non-)K$\ddot{\bf a}$hler manifolds
 and constrained dynamics}

\vskip 0.2cm

\vskip 0.5cm
{\bf Sergei V. SHABANOV}$\footnote{Alexander von Humboldt fellow}$

\vskip 0.5cm
{\em Institute for Theoretical Physics, Free University of Berlin\\
WE 2, Arnimallee 14, D-14195, Berlin, Germany}
\end{center}

\begin{abstract}
It is shown that $q$-deformed quantum mechanics
(systems with $q$-deformed Heisenberg commutation relations)
can be interpreted as an ordinary
quantum mechanics on K$\ddot{\rm a}$hler manifolds, or
as a quantum theory with second (or first)-class constraints.
\end{abstract}

{\bf 1}. The $q$-deformed Heisenberg-Weyl algebras \cite{1}, \cite{wor}
exhibiting the quantum group symmetries \cite{sk},\cite{drin} have attracted
much attention of physicists and mathematicians. In particular, some
works have been devoted to establishing potentials for ordinary quantum
systems which exhibit a $q$-deformed energy spectrum (see, for example,
\cite{bon} and references therein) in order to obtain classical dynamical
systems associated with the $q$-deformed Heisenberg-Weyl algebras.

In the present paper, we analyze the correspondence limit,
when $\hbar\rightarrow 0$ (rather than $q\rightarrow 1$),
of multimode $q$-deformed
Heisenberg-Weyl algebras \cite{1}, \cite{wor} regardless of the
specific form of a Hamiltonian. We show that classical
systems associated with the $q$-deformed Heisenberg-Weyl
algebras in this way
possess a non-trivial symplectic structure which, in turn,
is related to the  K$\ddot{\rm a}$hler symplectic structure
\cite{Beresin}. From the other hand, a natural (physical) reason
for a dynamical system to have a nontrivial symplectic structure
is the existence of "frozen" , non-dynamical degrees of freedom
or, in other words, constraints \cite{dir}. Dirac pointed out
that a nontrivial symplectic structure might naturally occur
in dynamical systems through
second-class constraints \cite{dir}. We shall
demonstrate below how constrained dynamics can be associated with
the $q$-deformed Heisenberg-Weyl algebra.

As was shown in \cite{2}, a
non-commutative phase space does not necessarily emerges in
the formal classical limit of $q$-deformed quantum mechanics, provided
the deformation parameter $q$ is
a function of the Planck constant. In the latter case,
$q$-deformed Heisenberg-Weyl algebras yield quadratic symplectic
structures on a commutative phase space.
For example, the one-mode $q$-deformed
Heisenberg-Weyl algebra defined by the commutation relation
\begin{equation}
\hat{b}\hat{b}^+-q^2\hat{b}^+\hat{b} =\hbar\ ,
\end{equation}
where $\hat{b}$ and $\hat{b}^+$ are creation and destruction operators,
turns into the following symplectic structure \cite{2}
\begin{equation}
\{b,b^*\}=-i(1-b^*b/\beta )\ ,
\end{equation}
where $\beta $ is a constant and $b$ and $b^*$ are commutative
complex coordinates on a phase-space plane.
Indeed, taking the classical limit
$[\hat{b},\hat{b}^+]/i\hbar=
-i(1-(1-q^2)\hat{b}\hat{b}^+/\hbar) \rightarrow \{b,b^*\}$
as $\hbar \rightarrow 0$, and $\hat{b},\hat{b}^+$
are simultaneously replaced by classical holomorphic
variables $b,b^*$, respectively, we see that Eq.(2) results from
(1) if one assumes $1-q^2=\hbar /\beta +O(\hbar^2)$.
Notice that without the latter assumption,
the formal limit $\hbar \rightarrow 0$ in (1) would lead to
a non-commutative phase space $bb^* =q^2b^*b$.

Yet, a canonical
quantization of the symplectic structure (2) ($[\ ,\ ]=i\hbar \{\ ,\ \}$)
leads to the deformed Heisenberg-Weyl algebra (1) \cite{3},\cite{2}.
The operator ordering ambiguity arising in the right-hand side of (2)
is irrelevant upon quantization. One can set, for example, $b^*b
\rightarrow \alpha\hat b^+\hat b + \alpha^*\hat b\hat b^+$ in the
right-hand side of (2), where
an arbitrary coefficient $\alpha$ obeys the correspondence rule,
$\alpha+\alpha^* = 1 +O(\hbar)$. Then the commutation relation (1)
is obtained by an appropriate renormalization of the creation and
destruction operators.

Hamiltonian equations of motion $\dot{b}=\{b,H\}$ depend on
a symplectic structure. If the Hamiltonian is
$H=\omega b^*b$
(a harmonic $q$-oscillator), then  $\dot{b}= -i(1-E/\beta)
b$, where $E=H$ is the oscillator energy, $\dot H =\{H,H\}=0$. Therefore, a
frequency of oscillations becomes a function of the oscillator energy.
Thus, one can regard a harmonic $q$-oscillator as a familiar anharmonic
oscillator \cite{2}.

{\bf 2}. Some other known
examples of the $q$-deformed Heisenberg algebra
($q$-particles) \cite{4}, \cite{5} can  also be treated as systems with
a "deformed" symplectic structure. Consider the following symplectic
structure
\begin{equation}
\{x,p\}=1+xp/\beta\ \ \ \ \ , \ \beta > 0\ ,
\end{equation}
where $x$ and $p$ are coordinate and momentum of a particle, respectively.
A canonical
quantization of (3) gives the $q$-deformed Heisenberg algebra discussed in
\cite{4}
\begin{equation}
\hat{p}\hat{x}-q\hat{x}\hat{p}=-i\hbar q^{1/2},\ \ \
\hat{p}^+=\hat{p},\ \ \ \hat{x}^+=\hat{x}\ ,
\end{equation}
with $q=e^{i\theta }$ and $\theta =\theta (\hbar /\beta )$ \cite{6}.
Indeed, by virtue of the canonical quantization rule we have from (3)
$[\hat{x},\hat{p}]=i\hbar (1+(\hat{x}\hat{p}+ \hat{x}\hat{p})/2\beta)$.
Relation (4) is obtained by renormalizing the operators $\hat{x}$ and
$\hat{p}$ with the coefficient $|1+i\hbar/2\beta|^{1/2}$, and
$q=(1-i\hbar/2\beta)/(1+i\hbar/2\beta)$.

The Hamiltonian equation of motion $\dot{x}=\{x,H\}$ of a free particle
$H=p^2/2$ induced by (3) coincides with the equation of motion
for a particle with friction.
The friction coefficient
depends on the deformation parameter $\beta $
and the particle generalized momentum $p$. Notice that $\dot{p}
=\{p,H\}=0$, therefore $p=p_0=const$, whereas $\dot x =
\{x,H\}= \gamma x + p_0,\ \gamma = p_0^2/\beta$.

A lattice quantum mechanics \cite{5},
\cite{7} appearing upon a deformation of the
Heisenberg algebra with a real $q$ \cite{5},
\begin{equation}
\hat{p}\hat{x}-q\hat{x}\hat{p}=-i\hbar,\ \
\hat{x}\hat{p}^+-q\hat{p}^+\hat{x}=i\hbar,\ \
\hat{p}^+\hat{p}=q\hat{p}\hat{p}^+,\ \ \hat{x}^+=\hat{x},
\end{equation}
can be obtained by quantizing a degenerate symplectic structure
\begin{equation}
\{x,p\}=1-ixp/\beta ,\ \ \{x,p^*\}=1+ixp^*/\beta ,\ \
\{p^*,p\}=ipp^*/\beta \ .
\end{equation}
The degeneracy ia due to the existence of an absolute integral of
motion
\begin{equation}
C=pp^*x/\beta -i(p-p^*)
\end{equation}
which commutes with all symplectic coordinates
$\{C,x\}=\{C,p\}=\{C,p^*\}=0$. Therefore, the system never leaves the
surface $C= const$ in due course. A phase space of the system is a
two-dimensional surface $C= const$. In quantum theory, eigenvalues
of the Casimir operator $\hat{C}$ determine irreducible representations of
the algebra (5) \cite{5}.

A straightforward application of the canonical quantization rule to (6)
encounters the same operator ordering problem as upon quantizing (2).
It must be resolved so that the Jacobi
identity is fulfilled on the quantum level
\cite{drin}, \cite{Beresin}. It is remarkable
that any operator ordering consistent with the Jacobi identity results in
the algebra (5). Different choices of the operator ordering correspond
to variations of terms $O(\hbar^2)$ in $q=q(\hbar,\beta)$ \cite{2}.

To get the quantum algebra
(5) from the classical Poisson algebra (6),
one can, for instance, postulate the first
commutation relation as follows $[\hat{x},\hat{p}] =
i\hbar(1-i\hat{x}\hat{p}/\beta)$, then $[\hat{x},\hat{p}^+]$ is obtained
by the Hermitian conjugation of the first one, assuming $\hat{x}^+=
\hat{x}$. The operator ordering in the last commutation relation in (6)
is fixed by the Jacobi identity,
 $[\hat x,[\hat p^+,\hat p]] + [\hat p^+,[\hat p,\hat x]] +
 [\hat p,[\hat x,\hat p^+]] =0$,
 and reads $[\hat{p}^+,\hat{p}]
=-\hbar\hat{p}\hat{p}^+/\beta$. So, $q=1-\hbar/\beta$.

Thus, the  $q$-deformed Heisenberg algebra can appear as a result of
quantizing a quad\-ra\-tic symplectic structure
\begin{equation}
\{\theta ^j,\theta ^k\}=
\stackrel{\circ}{\omega}^{jk}+c^{jk}_{in}\theta ^i\theta ^n
\end{equation}
where $\theta ^j$ is a set of real phase-space coordinates,
$\stackrel{\circ}{\omega}^{jk}$
is the canonical symplectic structure and $c^{jk}_{in}$
are "deformation" constants chosen so that the Jacobi identity for (8) is
satisfied.

Below we shall demonstrate that the symplectic structure resulting from the
$SU_q(n)$-covariant deformation of the Heisenberg-Weyl algebra is related
to a symplectic structure on K$\ddot{\rm a}$hler manifolds.

{\bf 3}. The following $q$-deformed commutation relations remains untouched
under the action of the quantum group $SU_q(n)$ \cite{wor}
\begin{eqnarray}
\hat{a}_i\hat{a}_j &= &q\hat{a}_j\hat{a}_i,\ \ \hat{a}_i^+\hat{a}_j^+=
\frac{1}{q}\hat{a}^+_j\hat{a}^+_i,\ \ i<j\ ;\\
\hat{a}_i\hat{a}_j^+&= &q\hat{a}^+_j\hat{a}_i,\ \ i\neq j\ ;\\
\hat{a}_i\hat{a}^+_i&-& q^2\hat{a}^+_i\hat{a}_i=\hbar
+(q^2-1)\sum\limits_{k<i}^{} \hat{a}_k^+\hat{a}_k\ .
\end{eqnarray}
To obtain a corresponding symplectic structure in a classical theory, one
may use the rule $[\ ,\ ]/i\hbar \rightarrow\{\ ,\ \}$ as $\hbar
\rightarrow 0$ and $\hat{a}_i,\ \hat{a}^+_j$  are simultaneously
to be changed by classical holomorphic variables $a_i,\ a^*_j$.
However, this is just a formal rule which sometimes helps
to guess a correct classical limit of a given quantum theory (see a
rigorous consideration in \cite{Beresin}). To make sure that this rule works
in the case of the algebra (9)-(11), we notice that by means of a
transformation proposed in \cite{cha} the commutation relations (9)-(11)
can be "diagonalized", be transformed to the form (1) for each oscillator
mode, while operators of different modes commute amongst each other. For
the commutation relation (1), the validity of the rule $[\ ,\ ]/i\hbar
\rightarrow \{\ ,\ \}$ can be rigorously
established in the framework of the path
integral formalism \cite{2}. Therefore the above mentioned formal
approach should give a correct classical mechanics in our case.
Assuming $1-q=\hbar /\beta +O(\hbar ^2)$
(otherwise there is no commutative phase space in the classical theory) we
arrive at the following Poisson bracket structure
\begin{eqnarray}
\{a_k,a_j\} &= &ia_ka_j/\beta ,\ \ \{a^*_k,a^*_j\}=-ia^*_ka^*_j/\beta ,
\ \ k<j\ ;\\
\{a_k,a^*_j\} &= &ia_ka^*_j/\beta ,\ \ \ k\neq j\ ;\\
\{a_j,a^*_j\}&=&-i\left(1-\frac{2}{\beta}\sum\limits_{k=1}^{j}
a^*_ka_k\right )\ ,
\end{eqnarray}
where $a_j,a^*_j$ are phase-space holomorphic coordinates.

Let us recall now a basic definition of the K$\ddot{\rm a}$hler manifold
\cite{Beresin}.
Let $z^i$ and $z^{k*}$ are complex coordinates on a manifold ${\cal M}$
and $g_{i\bar{k}}(z,z^*)$ is a metric tensor on it such that the
interval on ${\cal M}$ has the form $ds^2=g_{i\bar{k}}dz^idz^{k*}$ and
\begin{equation}
g_{i\bar{k}}=\pl ^2\phi /\pl z^i\pl z^{k*}\ ;
\end{equation}
the scalar function $\phi $ is called the K$\ddot{\rm a}$hler potential, and ${\cal M}$
is called
the K$\ddot{\rm a}$hler manifold. A K$\ddot{\rm a}$hler manifold turns into a symplectic
manifold if the following symplectic structure is introduced on it
\begin{equation}
\{A,B\}=ig^{\bar{j}k}\left(\frac{\pl A}{\pl z^k}\frac{\pl B}{\pl
z^{j*}}-\frac{\pl A}{\pl z^{j*}}\frac{\pl B}{\pl z^k}\right)
\end{equation}
for any two functions $A$ and $B$ of $z,\ z^*$, where $g^{\bar{k}i}$
is a matrix inverse to (15). The Poisson bracket thus defined obey the
Jacobi identity due to the property (15) \cite{Beresin}.

It is readily to see that the Poisson brackets (12)-(14) are not of the
K$\ddot{\rm a}$hlerian type because of (12). However, they can be
transformed to the form (16).
Indeed, the algebra (12)-(14) admits the following representation
\begin{equation}
a_i=z^i\prod\limits_{k=1}^{i-1}(1-2z^kz^{k*}/\beta )^{1/2}
\end{equation}
and $a^*_i$ is obtained by a complex conjugation of (17), where
\begin{equation}
\{z^j,z^{k*}\}=-i(1-2z^jz^{j*}/\beta )\delta ^{j\bar{k}}
\end{equation}
and $\{z^j,z^k\}=\{z^{k*},z^{j*}\}=0$.
This is, in fact, a classical analogy of the operator transformation
proposed in \cite{cha} to "diagonalize" the algebra (9)--(11).
Therefore, the K$\ddot{\rm a}$hler metric related to (18) reads
\begin{equation}
g_{i\bar {k}}=-\delta _{i\bar{k}}(1-2z^kz^{k*}/\beta )^{-1/2}\ .
\end{equation}
Representing the K$\ddot{\rm a}$hler potential in the form
\begin{equation}
\phi =\frac{\beta}{2}\sum\limits_{i}^{}
\varphi \left( \frac{2z^iz^{i*}}{\beta}
\right)
\end{equation}
and substituting (20) and (19) into (15) we obtain
\begin{equation}
\varphi (x)=-Li_2(x)=-\sum\limits_{k=1}^{\infty}\frac{x^k}{k^2}\ ,\ \
|x|<1\ ,
\end{equation}
with $Li_2$ being the Euler dilogarithm.

So, the $SU_q(n)$-covariant deformation
of the Heisenberg-Weyl algebra describes
a quantum theory on a K$\ddot{\rm a}$hler manifold with the potential (20),(21).

The phase space manifold with the metric (19) is curved. The scalar
curvature corresponding to the metric (19),
\begin{equation}
R=\sum_i\frac{8}{\beta}\left(1-\frac{2z^iz^{i*}}{\beta}\right)^{-1}\ ,
\end{equation}
tends to infinity as any of variables $z^i$ approaches the circle
$|z^i|^2=\beta/2$, assuming $\beta >0$. Therefore, for positive $\beta$
the phase space turns out to be compact, while for negative $\beta$ the
function (22) is regular on the entire complex plane.

{\bf 4}. Now we shall make a "bridge" between the $q$-deformation of the
Heisenberg-Weyl algebra and
constrained dynamics. It is known since long time ago that a non-trivial
symplectic structure may occur through second-class constraints \cite{dir}
in dynamical systems.
Let $\varphi _a(\xi)=0,\ \ a=1,2,\ldots ,2M$, are second-class
constraints on a phase space spanned by coordinates $\xi^i$,
i.e. the matrix $\{\varphi _a,\varphi _b\}=\Delta _{ab}$ is not
degenerate, where $\{\xi ^i,\xi ^j\}=\stackrel{\circ}\omega ^{ij}$ is the
canonical symplectic structure. Let $\xi ^i=\xi ^i(\theta )$ be a solution
of the constraints where physical variables $\theta ^\alpha ,\ \ \alpha
=1,2,\ldots ,2(N-M)$ are coordinates of a physical phase space,
$\varphi_a(\xi(\theta))\equiv 0$. Then a
symplectic structure on the physical phase space is induced by the Dirac
bracket \cite{dir}
\begin{equation}
\{A,B\}_D=\{A,B\}-\{A,\varphi _a\}\Delta^{ab}\{\varphi _b,B\}
\end{equation}
projected on the surface $\xi ^i=\xi ^i(\theta )$, here
$\Delta^{ab}\Delta_{bc}=\delta ^a_c$.

The induced symplectic structure might not coincide with the canonical
one, i.e. it might turn out be "deformed"
$\{\theta^i, \theta^j\}_D = \omega^{ij}(\theta)$.
One can raise a
question: is there such second-class constrained system
whose physical symplectic structure induced by the Dirac bracket (23)
has the quadratic form (8)? The answer is positive for the
simplest $q$-deformed systems considered in pp. 1 and 2 \cite{6}.
A generalization is rather simple.

Let $\omega ^{ij}(\theta )$ be a non-constant
symplectic structure and $\omega
_{ij}\omega ^{jk}=\delta ^k_j$. Let us
extend the initial phase space spanned
by $\theta ^j$ by adding new variables $\pi _j$ and postulate the canonical
symplectic structure on the extended phase space, $\{\theta ^j,\theta
^k\}=\{\pi _i,\pi _k\}=0$ and $\{\theta ^j,\pi _k\}=\delta ^j_k$, i.e. the
initial phase space serves as
a configuration space in the extended theory.
Following \cite{Bat} we introduce second class constraints as
\begin{equation}
\varphi _i(\pi ,\theta )=\pi _i+\bar{\omega }_{ij}(\theta )\theta ^j=0
\end{equation}
where
\begin{equation}
\bar{\omega }_{ij}(\theta )=(\theta ^i\frac{\pl}{\pl \theta
^i}+2)^{-1}\omega _{ij}(\theta )=\int\limits_{0}^{1} d\alpha \alpha\omega
_{ij}(\alpha \theta ) \ .
\end{equation}
Then \cite{Bat}
\begin{equation}
\{\theta ^i,\theta ^j\}_D=-\{\theta ^i,\varphi _k\}\Delta ^{kn}\{\varphi
_n,\theta ^j\}=\omega ^{ij}(\theta )\ .
\end{equation}
For the symplectic structures (18) or (12)-(14) the integral (25) can be
taken explicitly. Thus, $q$-deformed quantum mechanics may appear upon
quantization of a second-class constrained system.

{\em Remark}. Quantization of the quadratic symplectic
structure (8) (induced by the Dirac bracket (23)) is not obvious because
of the operation ordering. A naive application of the formal rule $[\ ,\
]=i\hbar \{\ ,\ \}$ can violet the Jacobi identity in quantum
theory (or associativity of the quantum algebra). One way to obtain an
associative quantum theory is to assume that the coefficients
$c^{kn}_{ij}$ are also to be changed
upon quantizing, $c^{kn}_{ij}\rightarrow
\tilde{c}^{kn}_{ij}(\hbar )$ so that $\tilde{c}^{kn}_{ij}(\hbar
=0)=c^{kn}_{ij}$ and $\tilde{c}^{kn}_{ij}$ provide associativity of
quantum theory. The latter yields some algebraic equations for
$\tilde{c}^{kn}_{ij}$ of the Yang-Baxter type to be solved.

Another way to manage the operator ordering problem is to convert the
second-class constrained dynamics (24) into the first-class ones with
sequent quantization \cite{Fad}, \cite{Bat} $\footnote{This procedure
is equivalent to quantization via the Darboux variables for a given
symplectic form. However, the goal of the conversion method is that
it allows us to avoid an explicit construction of the Darboux variables,
which is a rather hard problem in general.}$.
A curious observation in
this approach is that $q$-deformed commutation relations can also
occur through reducing a quantum first-class constrained (gauge) system to
physical (gauge-invariant) variables.
One should point out that a quadratic symplectic
structure is just a particular case in the framework of the conversion
method developed in \cite{Bat}.

{\bf 5}. For any symplectic matrix $\omega _{ij}(\theta )$ obeying the
Jacobi identity $\pl _k\omega _{ij}+{\rm cycle}(k,i,j)=0$, there exist
local Darboux coordinates in which the symplectic structure has the
canonical form \cite{Arnold}. Darboux variables for the Poisson bracket
(18) and, hence, for (12)-(14) (due to the relation (17)) can be
explicitly found \cite{2}.
Therefore, the quadratic "deformation" locally looks like a special
non-canonical transformation of the standard (Darboux) phase-space
coordinates \cite{2}. From the mathematical point of view, all phase-space
coordinate systems should be treated on equal footing. But in a physical
theory, phase-space coordinates are associated with observables, which
makes some particular coordinates dynamically distinguished. For
example, excitations of various physical systems can be modeled
through $q$-oscillators \cite{mas}, which means that all complicated
interactions in a physical system can be accumulated into $q$-deformed
commutation relations. Thus, $q$-deformed quantum mechanics can be
considered as an effective theory to describe physical excitations. It has
been, actually, illustrated in p.1 with examples of a $q$-oscillator and a
$q$-particle which are dynamically equivalent to an anharmonic oscillator
and a particle with friction, respectively.

In contrast with the above said, the interpretation within constrained
dynamics does not imply, in general, any non-trivial interaction leading
to "$q$-deformed" excitations. The $q$-deformation of the algebra of
observables may appear kinematically upon eliminating all unphysical
(gauge) degrees of freedom (i.e. after solving constraints).
\begin{center}
{\bf Acknowledgement}
\end{center}
The author is kindly grateful to Prof. I.A. Batalin for useful
discussions.

\end{document}